\documentstyle[prl,aps,psfig]{revtex}
\newcommand{\set}[2]{\newcommand{#1}{#2}}
\set{\im}{\mbox{Im\,}}
\set{\re}{\mbox{Re\,}}
\set{\veps}{\varepsilon}
\set{\cont}{\mbox{\scriptsize cont}}
\set{\tot}{\mbox{\scriptsize tot}}
\set{\free}{\mbox{\scriptsize free}}
\set{\corr}{\mbox{\scriptsize corr}}
\set{\bound}{\mbox{\scriptsize bound}}
\set{\scatt}{\mbox{\scriptsize scatt}}
\set{\bec}{\mbox{\scriptsize BEC}}
\set{\id}{\mbox{\scriptsize id}}
\begin{document}
\twocolumn[\hsize\textwidth\columnwidth\hsize
           \csname @twocolumnfalse\endcsname
\title{Medium modification of two-particle scattering in
nonideal Bose systems
}
\author{ H. Stein$^1$, K. Morawetz$^2$, and G. R\"opke$^{1,2}$}
\address{
$^1$ Fachbereich Physik, Universit\"at Rostock,\\
$^2$ MPG-AG "Theoretische Vielteilchenphysik" an der Universit\"at Rostock,\\
Universit\"atsplatz 1, D-18051 Rostock, Germany}
\maketitle
\begin{abstract}
Medium modification of scattering properties in interacting Bose 
systems are considered by solving the Bethe-Salpeter equation. 
An equation of state for the normal phase (generalized Beth-Uhlenbeck 
formula) is given
using the in-medium phase shifts to include two-particle correlations. 
Conclusions are drawn for systems of bosonic atoms with repulsive interaction 
such as sodium $^{23}$Na and rubidium $^{87}$Rb.
It is shown that the in-medium scattering length and the absolute value of 
the in-medium scattering phase shift for low scattering energies
increase with density.\\

\end{abstract}

\vskip2pc]
\section{Introduction}
Since the recent observation of Bose-Einstein condensation (BEC) in
rubidium $^{87}$Rb \cite{Cornell}, lithium $^{7}$Li \cite{Hulet} and 
sodium $^{23}$Na \cite{Ketterle} in atom traps at extremly 
low temperatures new interest has arisen in the investigations
of interacting Bose gases. For a comprehensive overview about the 
current experimental and theoretical activities see \cite{BEC93}.
The effect of finite size on the critical 
temperature of an ideal Bose gas, which is confined in a small cube, 
was investigated in \cite{GH95}. 
However, when performing experiments at higher 
densities, the question of the treatment of interaction in
a nonideal Bose gas becomes essential.

In \cite{S96} the temperature dependence of the moment of inertia of 
a dilute Bose gas in a harmonic trap and the role of interaction is discussed.
A common attempt to include interaction is the approximation of a
hard sphere Bose gas which can be treated analytically \cite{H}.
There, the interaction can be replaced by a pseudopotential containing 
the free scattering length $a_0$ as the only parameter.
For the atoms relevant in the recent BEC experiments, values for 
the scattering length are obtained in the low 
density limit from measurements of the cross section \cite{DMJAK95} 
or from evaluation of spectroscopic data \cite{{MV94},{GCMHBV95}}. 
However, the isolated two-particle scattering properties should be modified  
in dense systems due to the medium as shown below. This means that the
scattering processes in a dense medium cannot be interpreted in terms of
the free scattering length only. 

The medium modifications of two-particle properties result in different 
macroscopic effects. First, nonequilibrium processes are influenced. Usually, 
kinetic equations
are derived in RPA or Born approximation \cite{KD83,KD85,SW89}.
The corresponding collision terms include final state Bose enhancement. 
It was demonstrated \cite{SW89}, that approaching the critical density of
Bose condensation the scattering rates are increased due to Bose enhancement.
In addition, the influence of the medium can be included by a meanfield 
approximation in the framework of the quasiparticle concept. Employing 
the Bogoliubov transformation, a
quasiparticle approach was proposed by 
\cite{KD83,KD85}, which is applicable also below the critical temperature.

However, additional effects arise if we go beyond a quasiparticle Born 
approximation as shown below. These effects 
are obtained by the solution of the Bethe-Salpeter equation containing 
intermediate Bose enhancement.
They cause a modification of the $T$-matrix or the corresponding 
transition matrix elements  
which are absent in Born approximation. 
This $T$-matrix leads to medium dependend effective scattering phase shifts 
and 
cross sections which are an important ingredient for collision integrals.
In this paper we will not discuss kinetic theory in more detail (see 
\cite{MR94}).

Second, the thermodynamic quantities are modified by medium dependend 
two-particle properties. 
As well-known, nonideality effects of thermodynamic quantities 
can be expressed in terms of two-particle scattering phase shifts.
A relation between these phase shifts and the second virial coefficient can 
be established
using the Beth-Uhlenbeck formula \cite{BU36}.
It is possible to generalize the treatment of the isolated two-particle 
problem to dense quantum gases. This is achieved
using the $T$-matrix with intermediate Bose enhancement as we will show below.

Systematic attempts for the derivation of second virial corrections
for dense quantum gases are given using
thermodynamic Green function technique for Fermi
systems \cite{{ZS85},{SRS90}} or Ursell-operator technique \cite{GL95}.  
In this paper we apply thermodynamic Green functions for 
Bose systems and study the microscopic two-particle scattering 
problem in medium and the resulting equation of state. 
We discuss the conclusions
for two examples with repulsive interaction, sodium $^{23}$Na and 
rubidium $^{87}$Rb, which play a major
role in the BEC experiments.\\

First, let us remind some definitions and results of the 
scattering of two interacting particles in vacuum.
The free scattering length $a_0$ is defined in the limit of vanishing
relative momentum $k$  
\begin{equation}
a_0=-\lim_{k \to 0} \frac{\delta_0(k)}{k}\quad. 
\end{equation}
This follows from the phase shift $\delta_0(k)$ between
the asymptotic incoming and outgoing plane wave of the scattered particle,
which can be expanded for small relative momentum
\begin{equation}\label{dexp}
\cot \delta_0(k)=-\frac{1}{a_0k}+\frac{1}{2} kr_0  + O(k^3)\quad.
\end{equation}
The coefficient $r_0$ in the second term 
characterizes the range of the potential.\\
 
Let us consider an imperfect Bose gas with $a_0\ll\lambda$ where 
$\lambda=(\frac{2\pi \hbar^2}{mk_B T})^{1/2}$ is the thermal
wavelength.
In this case, a zero range pseudopotential can be employed which is 
parametrized by the free scattering length $a_0$ only.
For densities below the critical density of the ideal Bose-Einstein 
condensation $n_{\bec}=2.612 \lambda^{-3}$, a first order correction of
the ideal Bose gas pressure $p_{\id}$ can be derived \cite{H}
\begin{equation}\label{pHuang}
p(n,T)=p_{\rm id}(n,T)+\frac{4\pi \hbar^2 a_0}{m} n^2\quad. 
\end{equation}
In the nondegenerate limit $(n \ll n_{\bec})$, the equation of state 
(\ref{pHuang}) can be rewritten in 
terms of a virial expansion either of pressure or of density
\begin{eqnarray}\label{ve}
p(n,T)&=& nk_BT+B_p(T)n^2k_BT \nonumber\\ 
n(\mu,T)&=&n_{\rm id}(\mu,T)+B_n(T)n_{\rm id}^2(\mu,T) \quad,
\end{eqnarray}
with the second virial coefficient $B_p(T)=2a_0\lambda^2$ or 
$B_n(T)=-4a_0\lambda^2$.
In the following, we present a method of deriving the in-medium generalizations
of the scattering phase shifts and the scattering 
length as well as an equation of state
for a dense Bose gas with respective virial corrections in the normal phase.

\section{Green function approach to in-medium scattering and thermodynamics}

In a dense medium, the single-particle, two-particle, etc., energies
are modified compared to vacuum values. For instance, 
quasiparticle energies and exchange processes with the medium have to 
be included. A systematic quantum statistical approach can be given
employing thermodynamic Green functions. In the case of fermions, the 
in-medium scattering problem was investigated in 
\cite{{ZS85},{SRS90}}.  
We follow the method as outlined in \cite{SRS90} and apply it for the case of 
bosons.\\
We start from the Dyson equation  
\begin{equation}\label{dy}
G_1=G_1^0+G_1^0\Sigma G_1 \quad,
\end{equation}
which relates the free one-particle Green function 
$G_1^0(1;z)=(z-\veps_1)^{-1}$ and the 
the full one-particle Green function $G_1$ in a selfconsistent way.
The latter one can be written as
\begin{equation}\label{G1}
G_1(1;z)=(z-\veps_1-\Sigma(1;z))^{-1} \quad,
\end{equation}
where $1=\{p_1, \sigma_1\}$ denotes quantum numbers momentum and spin
and $\veps_1={\hbar^2 k_1^2\over 2m}$ is the free single particle energy.\\ 
The self energy $\Sigma$ can be expressed in terms of 
the $T$-matrix and the spectral function $A$ as
\begin{eqnarray}\label{se}
\Sigma(1;z)&=&\sum_2
\int_{-\infty}^{+\infty} \frac{dE_2}{2 \pi} A(2,E_2) 
\bigg[g_1(E_2)V_{ex}(12,12)\nonumber\\
&+&
\int_{-\infty}^{+\infty} \frac{dE}{\pi} \im T_{ex}(12,12;E)
\frac{g_1(E_2)-g_2(E)}{E-z-E_2}\bigg] \quad,\nonumber\\
\end{eqnarray}
where the index "$ex$" denotes the sum of direct and exchange matrix
elements in the respective quantity and 
$g_n(E)=\{\exp ((E-n\mu)/T)-1\}^{-1}$ 
is the $n$-particle Bose distribution function.

Using thermodynamic Green functions, the $T$-matrix of
the quantum scattering theory 
is generalized to finite temperature and density. 
It contains a partial summation of the interaction.
The resulting 
Bethe-Salpeter equation for the $T$-matrix within the ladder approximation
reads \cite{ZS85,SRS90,MR94}
\begin{eqnarray}\label{BSE}
T(12,1'2';z) &=& V(12,1'2')\nonumber\\
&+&\sum_{3456}V(12,34)\,
G_2^0(34,56;z)\,T(56,1'2';z).\nonumber\\
\end{eqnarray}
The $T$-matrix describes the two-particle properties and corresponds
to the Schr\"odinger equation but contains the influence of the medium
in the uncorrelated two-particle Green function $G_2^0$.
The uncorrelated two-boson Green function reads in spectral representation
\begin{eqnarray}\label{G2}
G_2^0(121'2',z)&=&\int_{-\infty}^{+\infty} \frac{dE_1}{2 \pi} 
              \int_{-\infty}^{+\infty} \frac{dE_2}{2 \pi}
\frac{1+g_1(E_1)+g_1(E_2)}{z-E_1-E_2}\nonumber\\
&&\nonumber\\
&&\times A(1,E_1) A(2,E_2) \delta_{11'} \delta_{22'}\quad.
\end{eqnarray}
The one-particle spectral function is defined as 
$A(1,E)=i [G_1(1,E+i0)-G_1(1,E-i0)]$ and
reads by using the Dyson equation (\ref{dy})
\begin{equation}\label{A}
A(1,E)=
\frac{2\, \im \Sigma(1,E)}
{[E-\veps_1-\re \Sigma(1,E)]^2+[\im \Sigma(1,E)]^2}\quad.
\end{equation}
The set of equations (\ref{se}), (\ref{BSE}), (\ref{G2}) and (\ref{A})
has to be solved selfconsistently.\\
 
The $T$-matrix is directly related to the two-particle scattering properties.  
E.g., in the continuum of scattering states the phase shifts $\delta$ 
can be expressed as
\begin{equation}\label{del}
\cot\delta (12,12;E)=\frac{\re T (12,12;E)}
                          {\im T (12,12;E)} 
\end{equation}
which contain the influence of the medium. These in-medium scattering 
phase shifts 
are a direct generalization of the standard definition of
free phase shifts which are included in the low density limit \cite{SRS90}.

From the spectral function thermodynamic properties of the
system can be derived. The one-particle density of a Bose system
reads 
\begin{equation}\label{n}
n(\mu,T)=\frac{1}{\Omega}\sum_{1} \int \frac{dE}{2 \pi} g_1(E) A(1,E) 
\end{equation} 
where $\Omega$ is a normalization volume.

\section{Extended quasi-particle approximation and 
generalized Beth-Uhlenbeck formula}

The full selfconsistent solution of eqs. (\ref{se}) to (\ref{A}) 
is a complicated task. Therefore, we consider the spectral function (\ref{A}) 
in quasi-particle approximation 
$A_{\rm qp}(1,E)=2\pi\delta(E-\veps_1-\Delta(1))$ 
which shows delta-peaked spectral weight at sharp quasi-particle energies 
$\epsilon_1=\veps_1+\Delta(1)$ of the single bosons.
The single-particle energy $\veps_1$ is shifted by
$\Delta(1)=\re \Sigma(1,z)|_{z=\epsilon_1}$ which is the real part of the
self energy (\ref{se}) at the quasi-particle energy.
The quasi-particle spectral function is applied to the self energy (\ref{se})
and to the uncorrelated two-boson Green 
function $G_2^0$ (\ref{G2}) which enters the ladder $T$-matrix (\ref{BSE}). 
From this a spectral function can
be obtained which goes beyond the quasi-particle approximation.
The so-called extended quasi-particle approximation is obtained by
assuming that the imaginary part of the self energy is small.
The resulting spectral function can be 
represented as a $\delta$-function and an additional 
contribution due to correlated particles 
in form of a derivation of the principal value \cite{{ZS85},{SRS90}}
\begin{eqnarray}\label{eqpA}
A_{\rm eqp}(1,E)&=&\frac{2\pi\delta(E-\epsilon_1)}
{1-{d\over dz}\re\Sigma(1,z)|_{z=\epsilon_1}} \nonumber\\
&-& 2\im\Sigma(1,E){d\over dE}
{{\rm P}\over E-\epsilon_1}
\end{eqnarray} 
Applying (\ref{eqpA}) in eq.(\ref{n}) results in an equation of 
state for the total density which contains
quantum statistical spectral weight from single
bosons in quasi-free or in possible two-particle bound and scattering states.
Similar to the classical equation of Beth and Uhlenbeck \cite{BU36}, the
total density splits into density contributions due to free
quasi-particles and correlated particles in bound and scattering states.
In an analogous way as done in \cite{SRS90} for a fermionic 
system we give a generalized Beth-Uhlenbeck formula
for the total density in a bosonic system which 
is only valid in the normal phase for temperatures above the
onset of Bose-Einstein condensation
\begin{equation}\label{gbu}
n_{\tot}(\mu,T)=n_{\free}(\mu,T)+2n_{\corr}(\mu,T)\quad, 
\end{equation}
with
\begin{eqnarray}
n_{\free}(\mu,T) &=& \frac{1}{\Omega}\sum_{1} g_1(\epsilon_1) \quad,\nonumber\\
n_{\corr}(\mu,T) &=& n_{\bound}(\mu,T)+n_{\scatt}(\mu,T) \nonumber\\
&=&\frac{1}{\Omega}\sum_{K} \sum_i g_2(E_{\cont}+E_i^b(K)) \nonumber\\
&+&\frac{1}{\Omega}\sum_{K} \int_{0}^{\infty} \frac{dE}{\pi}
g_2(E_{\cont}+E) \nonumber\\
&\times& \sum_{\alpha} c_{\alpha} \Bigl\{{d\over dE}
[\delta_{\alpha}(K;E)-{1\over 2}\sin(2\delta_{\alpha}(K;E))]\Bigr\}\nonumber
\end{eqnarray}
where, if necessary, summations over bound states and scattering
channels $\alpha$ with a corresponding degeneracy factor $c_{\alpha}$ 
have to be performed.
The in-medium bound state energy $E_i^b(K)$ depends on total momentum $K$,
chemical potential and temperature. 
Like the scattering energy $E$, 
it is taken relative to the energy at the continuum edge 
$E_{\cont}(K)=\hbar^2 K^2/4m+2\Delta(K)$ which
separates bound state ($E_i^b(K)<0$) and scattering state region ($E>0$).

Rewriting (\ref{gbu}) as
\begin{equation}\label{vgbu}
n_{\tot}(\mu,T)=n_{\free}(\mu,T)[1+B^*(\mu,T)\;n_{\free}(\mu,T)]
\end{equation}
for $(n_{\free}<n_{\bec})$, 
we can define a quantity $B^*(\mu,T)=2n_{\corr}(\mu,T)/(n_{\free}(\mu,T))^2$ 
which represents the non-ideality contribution in analogy to
the virial expansion of the total density (\ref{ve}).  
We remark that $B^*$ contains non-ideality contributions due to
two-particle interaction and degeneracy in any order of density,
but no three-particle interaction contributions. 
While the second virial coefficient $B_n(T)$ per definitionem contributes 
to second order in density and
depends on temperature only, $B^*$ also depends on chemical potential 
and hence on density.
In the low density and low temperature limit, 
or more exact for $a_0\ll \lambda$ and $n_{\free}\lambda^3\ll 1$, 
the non-ideality coefficient $B^*(\mu,T)$ 
coincides with the second virial coefficient $B_n(T)$ from eq. (\ref{ve}).

\section{Model calculation}
In the case of a separable representation of the interaction, 
the Bethe-Salpeter equation (\ref{BSE}) can be solved and the
$T$-matrix can be written in an analytic form.
The choice of a separable potential is motivated by the fact that 
there is a rigorous method to replace an arbitrary 
potential by a rank-$N$ separable
potential \cite{EST}.
Separable potentials are applied in the description 
of different physical systems, e.g., by Nozi\`eres and Schmitt-Rink
\cite{NS85} for electron-hole interaction or in nuclear physics
for the nucleon-nucleon interaction \cite{Y54}. As shown below
the contact interaction which is widely used in atomic physics is
a special case of a separable potential.

For our model calculation we choose a rank-one separable potential of the form
\begin{equation}\label{V}
V(12,1'2')=V(kK,k'K')  =   V_0\, v(k)v(k')\delta_{KK'}\quad,
\end{equation}
with a formfactor
\begin{equation}
\quad v(k) = \frac{1}{k^2+\beta^2} \quad.\nonumber
\end{equation}
The potential depends on the square of the incoming and outgoing relative 
wave vector $k=|\vec k_1-\vec k_2|/2$ and 
$k'=|\vec k_1'-\vec k_2'|/2$ and the two parameters $\beta$ 
($\approx$ inverse range) and $V_0$ (strength), but it is
independent of the center-of-mass momentum $\hbar K$.
For $\beta\to\infty$ the special case of a contact potential, i.e. 
a zero range interaction, is included.

Now, using relative coordinates, we can solve the in-medium Bethe-Salpeter
equation (\ref{BSE}) explicitly
\begin{equation}\label{st}
T(kK,k'K';z)=\frac{V(kK,k'K')}{1-J(K;z)}\quad,
\end{equation}
with
\begin{equation}
J(K;z)=\sum_{k',K'} V(k'K,k'K')\,G^0_2(k'K;z)\quad.\nonumber
\end{equation}
The two-boson Green function in quasi-particle approximation reads
\begin{equation}\label{G20}
G_2^0(kK;z)=\frac{Q(K,k)}{z-\epsilon_1-\epsilon_2}\quad.
\end{equation}
The medium effects enter via the Bose enhancement factor $Q(K,k)$
and via the self energy shift of the quasi-particle energies.
For our exploratory calculation below we 
replace the shift by its thermal average $\bar\Delta$. This rigid shift 
can be incorporated in an effective chemical potential 
$\mu^*=\mu-\bar\Delta$.

The integration of angles in the Bose distributions
$Q(K,k)=\int\frac{d\Omega}{4\pi}\left[1+g_1(\epsilon_1)+g_1(\epsilon_2)\right]$
gives an analytic expression for the Bose enhancement factor
\begin{equation}\label{Q}
Q(K,k)=1+{2 m k_B T \over \hbar^2 K k} {\rm ln}\left |
{1-{\rm exp}\left (-{\hbar^2({K \over 2}+k)^2 \over 2 m k_B T}+
 {\mu^* \over k_B T}\right)
 \over
 1-{\rm exp}\left (-{\hbar^2({K \over 2}-k)^2 \over 2 m k_B T}+
 {\mu^* \over k_B T}\right)
}\right | \quad.
\end{equation}
Now, the quantity $J$ reads explicitly
\begin{equation}\label{J}
J(K,E(k)+i0)={V_0\over 2\pi^2} \int\limits_0^{\infty} dk'\; k'^2 
{v(k')^2\;Q(K,k') \over E(k)-{\hbar^2 k'^2\over m} +i0} \quad.
\end{equation}
Here, the scattering energy $E(k)$ is taken relative 
to the energy at the continuum edge $E_{\cont}(K)$.
The in-medium scattering phase shift follows
according to eq.(\ref{del}) and eq.(\ref{st})
\begin{equation}\label{delJ}
\cot\delta (K;E(k))
=\frac{1-\re J (K;E_{\cont}(K)+E(k))}
      {  \im J (K;E_{\cont}(K)+E(k))} \quad.
\end{equation}
The small $k$ expansion (\ref{dexp}) can be applied to the in-medium 
phase shifts $\delta$ in the
same way as for the free phase shift and provides now a 
quantity which is a direct generalization of the free scattering 
length $a_0$. In analogy to the phase shift it is named
in-medium scattering length $a(\mu,T)$ and follows in the center-of-mass system 
($K=0$) from 
\begin{eqnarray}\label{a}
a(\mu,T)&=&-\lim_{k \to 0}\; \frac{\im J(K,E_{\cont}(0)+E(k))}
                    {k\; [1-\re J(K,E_{\cont}(0)+E(k))]}\quad.
\end{eqnarray}
If we consider a repulsive potential (no bound states) and
cooled spin-polarized atoms (triplet scattering only) as used in magnetic traps,
the generalized Beth-Uhlenbeck formula eq.(\ref{gbu}) simplifies 
after partial integration to
\begin{eqnarray}\label{bbu}
\lefteqn{n_{\tot}(\mu,T) = {1\over\Omega}\sum_{1} g_1(\epsilon_1) 
+2\int_0^{\infty} \frac{dK\,K^2}{2\pi^2}\int_0^{\infty} 
\frac{dE}{\pi}}\nonumber\\
&\times& \frac{\delta(K;E)}
{2k_BT [\cosh ((E+E_{\cont}(K)-2\mu^*)/k_BT)-1]} \;.
\end{eqnarray}\\

Now we focus on the case of a contact potential. 
This is done to compare with classical results as eq.(\ref{ve})
which are also obtained with a zero range pseudopotential. A physical
reason is the assumption $r_0\ll a_0\ll \lambda$ which makes the first term in
expansion (\ref{dexp}) the dominant one. It is valid for the atomic systems
considered below if we have low temperatures and a potential range $r_0$
in the order of the atomic hard core.
  
For a zero range potential ($\beta\to\infty$ in eq.(\ref{V})) 
the $T$-matrix takes the form
\begin{eqnarray}\label{TCI}
T&\Bigl(&K;E={\hbar^2k^2\over m}\Bigr)={4\pi\hbar^2 \; a_0\over m}\nonumber\\
&\times& \Bigl(1+ia_0 k
+{a_0\over\pi}\int\limits_0^{\infty} dy 
{\sqrt{y}\over y-k^2} (Q(K,\sqrt{y})-1)\Bigr)^{-1} \,.
\end{eqnarray}
Here, the free $T$-matrix is first integrated out and then 
the limit $\beta\to\infty$ is taken \cite{R94-392}. This procedure 
avoids the introduction of a cut-off parameter to regularize the
integration. 

The in-medium scattering phase shift for the zero range interaction
reads 
\begin{eqnarray}\label{dCI}
\delta&\Bigl(& K;E={\hbar^2k^2\over m}\Bigr)=\nonumber\\
&-& {\rm arctan}\Biggl({ a_0 k\;Q(K,k)\over
1+{a_0\over\pi}\int\limits_0^{\infty} dy 
{\mbox{P}\over y-k^2} \sqrt{y}(Q(K,\sqrt{y})-1)}\Biggr)\,.
\end{eqnarray}
It differs from the free scattering phase, which is simply 
$\delta_0(k)=-{\rm arctan}(a_0 k)$, by the occurrence of the Bose enhancement
factor $Q$ according to eq. (\ref{Q}) in the nominator and in the principal
value integration in the denominator. 
Then, the in-medium scattering length for the zero range interaction is
\begin{equation}\label{aCI}
a(\mu,T)=a_0\Biggl({1+2g_1(0)\over
1+{4a_0\over\pi}\int\limits_0^{\infty} dk'g_1(\epsilon_{k'})}\Biggr)\quad.
\end{equation}
The free scattering length $a_0$ is modified by the
Bose functions in the nominator and denominator.  
For $\mu^*\to 0$ the Bose pole in the nominator dominates the
expression (\ref{aCI}) and the in-medium scattering length diverges
as one approaches the onset of Bose-Einstein condensation.\\

To apply the considerations above we focus on two systems which
play an important role in the current discussion on atomic Bose gases. 
For the first one,
sodium $^{23}$Na, a free ground state 
scattering length of $\pm (92\pm 25) a_B$ (in units of the  Bohr radius $a_B$)
is obtained from cross section measurement \cite{DMJAK95}.  
Spectroscopic data are in agreement with
$64<a_0/a_B<152$ \cite{MV94}, i.e. repulsive interaction is favoured
and bound state formation is excluded.
For the second example, rubidium $^{87}$Rb, the
ground state triplet scattering length $a_0\approx 5$~nm is of the same order 
as for sodium \cite{Cornell}.
Spectra are analyzed to yield values $85<a_0/a_B<140$ \cite{GCMHBV95}.
For our exploratory calculation we have chosen a free scattering 
length $a_0=92~a_B$ for sodium and $a_0=95~a_B$ for rubidium $^{87}$Rb to
parametrize the $T$-matrix (\ref{TCI}).

\section{Results and discussion}
\begin{figure}[htbc]
\centerline{\psfig{figure=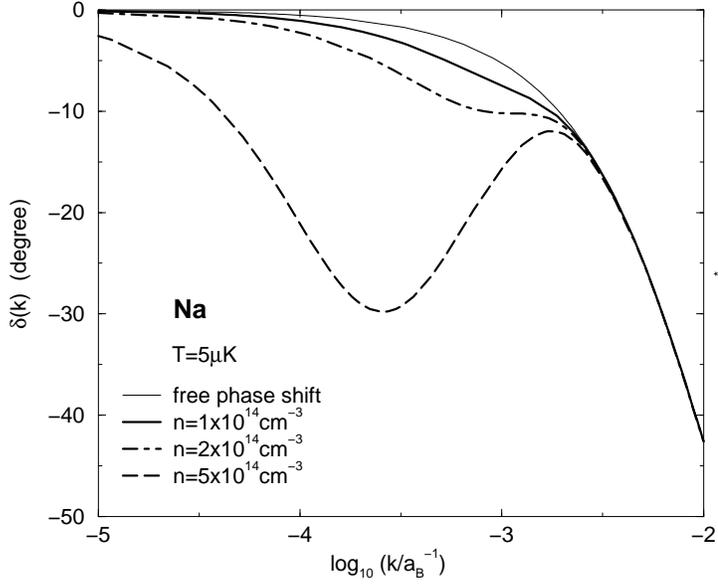,height=9cm}} 
\caption{Free and in-medium scattering phase shift  
versus decadic logarithm of relative momentum for sodium. 
Temperature is $T=5\mu$K.
\label{1}}
\end{figure}

\begin{figure}[htbc]
\centerline{\psfig{figure=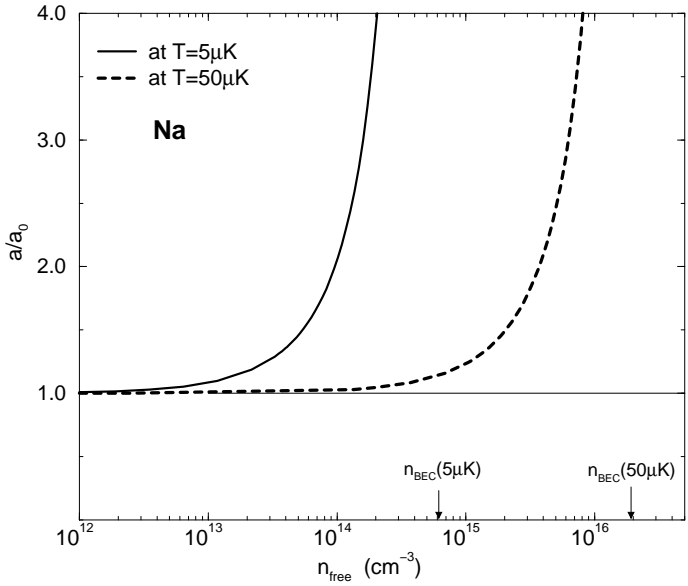,height=9cm}} 
\caption{The in-medium scattering length $a(\mu,T)$ in units of 
the free scattering length $a_0$ 
versus free density for sodium at temperatures $T=5\mu$K and $T=50\mu$K.
The critical densities are marked. 
\label{2}}
\end{figure}

\begin{figure}[htbc]
\centerline{\psfig{figure=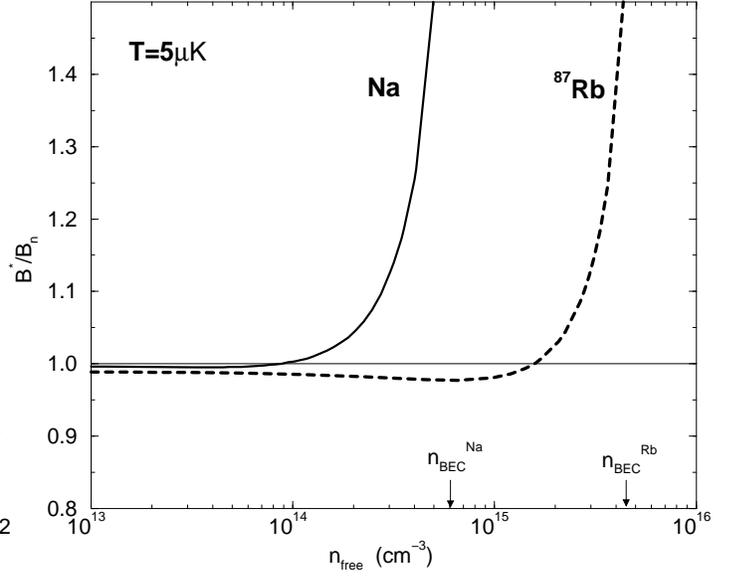,height=9cm}} 
\caption{
The nonideality coefficient $B^*(\mu,T)$ in units of the
classical second virial coefficient
$B_n(T)$ for sodium and rubidium versus free density $n_{\free}(\mu,T)$
at temperature $T=5\mu$K.
\label{3}}
\end{figure}

Figure \ref{1} shows the result for the in-medium scattering phase shift of 
sodium according to eq.(\ref{dCI}). For some densities the dependence 
on relative momentum  
is plotted and compared to the free scattering phase shift. 
The temperature $T=5\mu$K has been
chosen to have similar values as reached in the recent experiments. 
E.g., in the sodium experiment \cite{Ketterle} densities
exceeding $10^{14} {\rm cm^{-3}}$ have been produced and a
condensate has been observed at temperatures below $2~\mu$K.
The densities in our calculation are below the
critical density of $n_{\bec}=6\times 10^{14}{\rm cm^{-3}}$ at $T=5\mu$K.
In general, medium modifications of the phase shift
compared to the free scattering phase shift become small ($<1\%$)   
for scattering energies $\hbar^2 k^2/m > (10...20)k_BT$. 
The influence of the medium enlarges the absolut value
of the phase shift and leads to an extra minimum if one 
increases the density towards the critical value.
Furthermore, we see that the medium increases the
derivative of the phase shift at small momenta. This will result
in an increase of the scattering length as 
demonstrated in figure \ref{2}.
Since the free scattering length appears
to be of the same size for sodium and rubidium 
the major difference in the results
originates from the different atomic mass. Consequently,
compared to sodium at the same temperature 
a similar behaviour of scattering shift and length occurs for rubidium 
at densities about seven to eight times higher.\\
Figure \ref{2} shows that the scattering length according to eq.(\ref{aCI}) 
enlarges with increasing density due to the Bose enhancement.
We have plotted the density
dependence of the in-medium scattering length of sodium
for two different temperatures. It shows a strong increase if the
density approaches the critical density $n_{\bec}$.  
According to our model calculation for sodium the value of the
scattering length is doubled for $T=5\mu$K at a density of 
$1\times 10^{14}{\rm cm^{-3}}$ ($n_{\bec}=6\times 10^{14}{\rm cm^{-3}}$)
and for $T=50\mu$K at a density of
$3.7\times 10^{15}{\rm cm^{-3}}$ ($n_{\bec}=1.9\times 10^{16}{\rm cm^{-3}}$). 
For rubidium $^{87}$Rb at $T=5\mu$K the
scattering length is doubled at $8\times 10^{14}{\rm cm^{-3}}$
($n_{\bec}=4.5\times 10^{15}{\rm cm^{-3}}$).
If one measures the cross section 
a clear enlargement of the scattering length 
should be observed already at densities about 5 times smaller
than the critical density for Bose-Einstein condensation. \\
In figure \ref{3} we have evaluated the generalized Beth-Uhlenbeck
equation as given in eq.(\ref{bbu}) and have compared
the non-ideality coefficient $B^*(\mu,T)$ of eq.(\ref{vgbu})
with the classical second virial coefficient $B_n(T)$ of eq.(\ref{ve}).  
The relation of both is plotted for $^{23}$Na and $^{87}$Rb versus the 
free density of quasi-bosons at a fixed temperature $T=5\mu$K.
Approaching the critical density, the non-ideality coefficient $B^*(\mu,T)$
increases strongly.
The mass difference causes the respective behaviour for
rubidium to occur at roughly seven times higher densities then for sodium
at this temperature.
In the low density limit $B^*$ approaches the second virial coefficient
$B_n$. Figure \ref{3} shows a minor deviation especially for $^{87}$Rb
from the expected low density limit $B^*/B_n=1$. 
It can be traced back to the limited validity of
the condition $a_0\ll\lambda$ necessary to derive (\ref{pHuang}). 
The classical second virial coefficient $B_n$ as the low-density limit 
is slightly overestimated for
higher temperatures and greater masses. We stress that the absolute
value of the non-ideality coefficient $B^*(\mu,T)$ is monotonically
increasing with density.

We have also performed calculations with a separable potential of
finite range. The model parameter $V_0$ and $\beta$ of
the potential (\ref{V}) have been fitted to the free scattering 
length $a_0$ and the
potential range $r_0$. The latter one characterizes the short range
behaviour of the interaction. Since no exact datas are available,
we have taken a value in the order of the 
hard core radius of the atoms.  
A theoretical value of the hard core radius  
is about 4$a_B$ (from \cite{RR89} at higher temperatures).
The results with this finite range interaction 
have been found to be not sensitive if compared 
to the results presented in the figures 1-3 for zero range interaction.
This is not very surprising since $r_0$  
is much smaller than the free
scattering length $a_0$. But we are cautious to state 
that the results are in general independend of the 
potential range since we know neither $r_0$ nor a hard core radius at this
very low temperatures.  

We remind that we restricted our calculations to the normal phase.
Improving the ordinary quasiparticle picture, correlations are included
within the so-called extended quasiparticle approximation.
In the Bose-condensed phase the common approaches within 
Hartree-Fock-Bogoliubov approximation (see \cite{{KD83},{G96}}) 
should be improved on the same level by including correlations
beyond meanfield.
 
Furthermore, our calculations are restricted to the 
case of repulsive
interacting bosons. The consideration of an attractive interaction
leads to the possibility of an additional liquid-gas phase transition
and possibly to the formation of bound states. The
question of a superposition of a gas-liquid or gas-solid phase
transition and the Bose-Einstein condensation in an attractive atomic Bose gas
like lithium $^7$Li is discussed in \cite{S94}. There, the equation of state 
is extended to the Bose-condensed region.
In the normal phase it
is considered in the approximation of eq.(\ref{pHuang}) which does not include
in-medium scattering and possible
contributions from bound states. A treatment in the framework of 
the generalized Beth-Uhlenbeck equation (\ref{gbu}) would provide
a better understanding of the possible phase transitions.

In a generalization of the derived results to non-equilibrium processes,
transport coefficients should be modified not only due to final state Bose 
enhancement \cite{{KD85},{SW89}}, but as well due to intrinsic medium effects.
The ladder $T$-matrix should serve as an ingredient for more advanced 
formulations of kinetic equations including strong correlations \cite{MR94}.
  
In conclusion we suggest that in the recent experiments with Bose gases, 
such as sodium and rubidium in atom traps, density and
temperature regions are within reach where 
the scattering properties and the thermodynamic properties are 
influenced by the medium. We expect that the medium modifications
have direct impact on the experimentally extracted values of
the cross section and hence the scattering length.\\

\begin{center}{\bf Acknowledgements}\\
\end{center}

We thank Franck Lalo\"e and Peter Gr\"uter for stimulating discussions.

\newpage

\begin{thebibliography}{10}

\bibitem{Cornell}
M.H. Anderson, J.R. Ensher, M.R. Matthews, C.E. Wieman, E.A. Cornell,
Science {\bf 269},  198  (1995).

\bibitem{Hulet}
C.C. Bradley, C.A. Sackett, J.J. Tollett, and R.G. Hulet, 
Phys. Rev. Lett. {\bf 75},  1687  (1995).

\bibitem{Ketterle}
K.B. Davis, M.-O. Mewes, M.R. Andrews, N.J. van Druten, D.S. Durfee,
D.M. Kurn, and W. Ketterle, Phys. Rev. Lett. {\bf 75},  3969  (1995).

\bibitem{BEC93}
{\it Bose-Einstein Condensation}, ed. by A. Griffin, D.W. Snoke, and S. 
Stringari, Cambrigde University Press, Cambridge (1995).

\bibitem{KD83}
T. R. Kirkpatrick, J. R. Dorfman, Phys. Rev. A {\bf 28}, 2576 (1983).

\bibitem{KD85}
T. R. Kirkpatrick, J. R. Dorfman, J. Low Temp. Phys. {\bf 59}, 1 (1985).

\bibitem{SW89}
D. W. Snoke, J. P. Wolfe, Phys. Rev. B {\bf 39}, 4030 (1989).

\bibitem{GH95}
S. Gro{\ss}mann and M. Holthaus, Z. Naturforsch. {\bf 50a}, 323 (1995).

\bibitem{S96}
S. Stringari, Phys. Rev. Lett. {\bf 76},  1405  (1996).

\bibitem{H}
K. Huang, {\it Statistical Mechanics}, J. Wiley, New York (1987).

\bibitem{DMJAK95}
K.B. Davis, M.-O. Mewes, M.A. Joffe, M.R. Andrews, and W. Ketterle, 
Phys. Rev. Lett. {\bf 74},  5202 (1995).

\bibitem{MV94}
A.J. Moerdijk and B.J. Verhaar, Phys. Rev. Lett. {\bf 73},  518  (1994).

\bibitem{GCMHBV95}
J.R. Gardner, R.A. Cline, J.D. Miller, D.J. Heinzen, H.M.J.M.
Boesten, and B.J. Verhaar, Phys. Rev. Lett. {\bf 74},  3764  (1995).

\bibitem{BU36}
G.E. Uhlenbeck and E. Beth, Physica {\bf 3},  729  (1936);
E. Beth and G.E. Uhlenbeck, Physica {\bf 4},  915  (1937).

\bibitem{ZS85}
R. Zimmermann and H. Stolz, Phys. Status Solidi B {\bf 131},  151  (1985).

\bibitem{SRS90}
M. Schmidt, G. R{\"o}pke, and H. Schulz, Ann. Phys. (NY) {\bf 202},  57
(1990).

\bibitem{GL95}
P. Gr\"uter, F. Lalo\"e, J. Phys. I France {\bf 5} 181(part I) (1995),
1255(part II) (1995).

\bibitem{MR94}
K. Morawetz and G. R{\"o}pke, Phys. Rev. E {\bf 51},  4246  (1995).

\bibitem{EST}
D.J. Ernst, C.M. Shakin, and R.M. Thaler, Phys. Rev. C {\bf 8},  46 (1973).

\bibitem{NS85}
P. Nozi\`eres, S. Schmitt-Rink, J. Low Temp. Phys. {\bf 59} 195 (1985).

\bibitem{Y54}
Y. Yamaguchi, Phys. Rev. {\bf 95},  1628  (1954).

\bibitem{R94-392}
G. R\"opke, JINR Dubna preprint, E17-94-392 (1994).

\bibitem{RR89}
R. Redmer and G. R\"opke, Contrib. Plasma Phys. {\bf 29}, 343 (1989).

\bibitem{G96}
A. Griffin, Phys. Rev. B {\bf 53}, 9341 (1996).

\bibitem{S94}
H.T.C. Stoof, Phys. Rev. A {\bf 49}, 3824 (1994).

\end{thebibliography}
\end{document}